\documentclass[12pt,a4paper]{article}
\usepackage[margin= 1 in]{geometry}
\usepackage{graphicx}
\usepackage{latexsym}
\usepackage{amsmath}
\usepackage{scalerel}
\usepackage[utf8]{inputenc}
\usepackage[T1]{fontenc}
\usepackage{tcolorbox}

\title{Radiation from a Classical Harmonic Oscillator}
\author{Paritosh Verma \\ Narodowe Centrum Bada$\acute{n}$  J\c{a}drowych \\ Andrzeja Sołtana 7, 05-400 Otwock, Polska \\ Paritosh.Verma@ncbj.gov.pl}
\date{}

\begin{document}

\maketitle

\section{Abstract}

This article presents the calculation of gravitational and electromagnetic radiation  emitted from a classical simple harmonic oscillator (SHO). Here we show only the selected formulae and apply them to a toy problem without rigorous derivation. First, we calculate the explicit expressions for the gravitational waves polarizations and then obtain the power radiated away in gravitational waves using Brans-Dicke theory (BD). We also calculate the electromagnetic dipole and quadrupole power emitted by the SHO and compare them with their gravitational counterparts. 


\section{Introduction }

n 1961, Robert H. Dicke and Carl H. Brans proposed a scalar-tensor theory, known as BD theory \cite{Brans-Dicke}, to describe gravitation by incorporating Mach's principle. The foundation of this theory was built on the previous work of Pascual Jordan \cite{Jordan}  as well as Markus Fierz \cite{Fierz}, and sometimes it is also referred to as Jordan-Fierz-Brans-Dicke theory. In Einstein's general relativity (GR), the coupling constant between matter and spacetime is given by G (the Newtonian constant of gravitation).  In BD theory, the coupling depends also on a parameter $\zeta$ through the relation $G (1 - \zeta) $. The BD parameter $(\zeta)$ is obtained through experiments, and the Cassini \cite{Bertotti-Tartora} experiment in 2003 has imposed the constraint $ \zeta < 0.0000125 $.

A generic metric theory of gravity can possess up to six polarization states of gravitational waves (GWs); two tensor states; two vector states and two scalar states \cite{Isi}. GR has only two tensor polarizations, whereas BD has three polarizations. The first two are tensor polarizations similar to GR, and the third is scalar polarization. In recent years, there has been a decent amount of work either to test GR using the LIGO-Virgo-Kagra (LVK) data \cite{GR test GWTC3}, \cite{GR-test-GW170817}  or to modify gravity to explain cosmological ambiguities like the nature of dark energy and Hubble tension \cite{Bertolami}, \cite{JBD-cosmo}, \cite{Sola}. There are many promising theories, but we choose BD for the following reason. BD theory is not too far from GR because the tensor polarizations differ from the ones in GR only by a tiny parameter $\zeta$. However, with the current sensitivity of detectors, it would be too hard to differentiate between the tensor polarizations of BD and GR. And we don?t want any new theory to deviate much from Einstein?s theory because it has already passed many tests. So, BD theory appears to be a good candidate to start with. Recently, we have implemented the BD theory to hunt for the dipole radiation in the LVK O3 data \cite{BD-LVK} by adding the D-statistics developed in \cite{Verma}.

We consider a simple harmonic oscillator moving in the x-direction, as shown in Figure 1. The spring is ideal and massless. We isolate the system so that the only force acting on the mass m is the spring force. The point particle also carries a charge q and the equation of the oscillator is given by 

 \begin{equation}\label{eq:0.0}
x_p(t) = x_0 \cos \omega t
 \end{equation}

where $\omega^2 = \frac{k_{eff}}{m} $ and $k_{eff}$ is the effective spring constant. Since, we are dealing with a classical SHO, we assume that  $x_0 \omega << c$ where $c$ is the speed of light.

This problem is a simple mechanical problem that provides insight into radiation without any knowledge of astrophysics. Also, we start with the BD theory and then take a limiting case which leads to GR.

\begin{figure}[!t]
\caption{A point particle with mass m and carrying a charge q is connected to a massless ideal spring. The other end of the spring is connected to a fixed wall. The particle can oscillate in the x-direction about the mean position O.}\label{sho}

\vskip -12pt
\centering
\includegraphics[width=6.5cm, height=6.5cm]{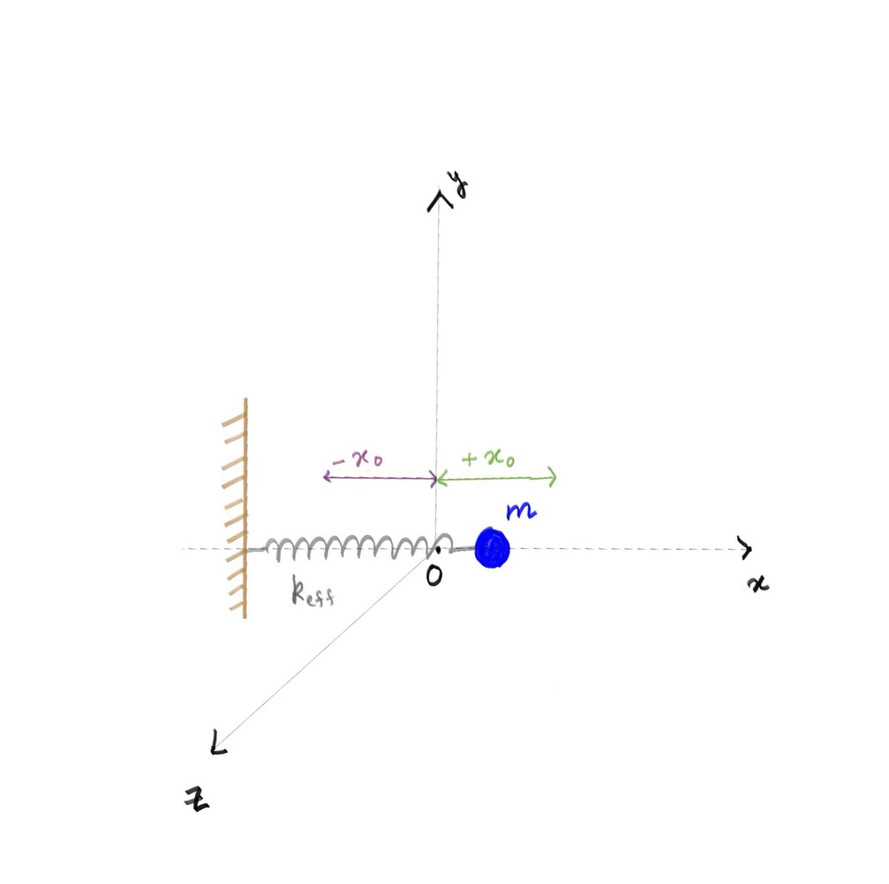}
\end{figure}



\section{A theoretical minimum on radiation}

\subsection{Gravitational radiation}

This subsection briefly presents the formulae for GW polarizations in BD theory and the power radiated from the system. There are three polarization states in BD theory; two tensor polarizations and one scalar polarization.  In the field theory language, the tensorial states could be conceived as spin-2 particles, whereas the scalar one as a spin-0 particle, and they all are massless because gravity is a long-range force. The two tensor polarizations are transverse to the wave, and they differ from each other by 45 degrees. So it means that if we rotate one polarization by 45 degrees, we get to the second polarization. And a rotation of 180 degrees gets back to the same state. A circle can represent the scalar polarization in the BD theory. This means that a rotation by an infinitesimally small angle leads to the same polarization.

The difference in polarization can also be understood in the language of particle physics. Let $\phi$ be the minimum angle by which one polarization should be rotated to get the same state. The $\phi$ is given by

 \begin{equation}\label{eq:0}
\phi = \frac{2 \pi^c}{s}
 \end{equation} 
 
 where s is the spin of the graviton. For scalar polarization, $s=0$, $s=1$ for vector polarization and $s =2$ corresponds to the tensor polarization. This formula is valid only for Bosons (integer spin particles) and not for Fermions (half odd integer spin particles). If we substitute $s=0$ in the above formula, $\phi$ tends to infinity, which implies that even an infinitesimal rotation leads to the same polarization.  We can call tensor polarizations $2$-fold symmetric, vector polarizations $1$-fold symmetric and scalar polarization $\infty$-fold symmetric. 

Here, we present only important formulae without any derivations. For a detailed analysis, one can follow  \cite{Verma}, \cite{Krolak}.

Assuming that the wave propagates in the $+z$ direction,  we have the following formulae for $h_{+}$  and $h_{\times}$ polarizations

\begin{eqnarray} 
\label{eq:1}
h_{+} (t) &=& \frac{G}{r c^{4}} (1 - \zeta) 
(\ddot{Q}^{xx}_{mW} (t^{\prime}) - \ddot{Q}^{yy}_{mW} (t^{\prime})) 
\nonumber \\
h_{\times} (t) &=& \frac{2 G}{r c^{4}} (1 - \zeta)  \ddot{Q}^{xy}_{mW} (t^{\prime})
\end{eqnarray} 

where G is the gravitational constant, c is the speed of light, $t^{\prime}$ is the retarded time, r is the distance of the source, $\ddot{Q}^{ij}_{mW} $ is the second time derivative of the mass quadrupole moment in the wave frame. In the limit $\zeta  \rightarrow  0$ the above expressions for the two polarizations reduce to the expression in classical general relativity given by Eqs.\, (1.114) of \cite{Krolak}.
 
 The scalar polarization $(h_S)$ in BD theory is given by 
 
\begin{equation} \label{eq:1.1}
h_{S}(t) = \frac{2G}{r c^{2}}  \zeta \left[  M(t^{\prime})  +  \frac{1}{c} \dot{D}_{mW}^{z} (t^{\prime})  - \frac{1}{2 c^2}   \ddot{Q}^{zz}_{mW}(t^{\prime})        \right]
 \end{equation}

where $\dot{D}_{mW}^{i}$ is the first time derivative of the mass dipole moment in the wave frame and $M$ is the mass monopole moment. In GR, there is no contribution from the mass monopole moment and dipole moment because of mass and linear momentum conservation. But monopole and dipole radiation appear in BD theory because the scalar field in BD theory does not satisfy a conservation law.

The total power radiated away per unit angle in BD theory can be written as

\begin{equation} \label{eq:2}
  \frac{dP_{grav}}{dA} = \frac{c^3}{16 \pi G (1 - \zeta) } <   \dot{h}^2_{+}(t) + \dot{h}^2_{\times}(t) +  \left( \frac{1 - \zeta}{\zeta} \right)  \dot{h}_{s}^2(t)   >  = \frac{dP^{(T)}}{dA} + \frac{dP^{(S)}}{dA}
\end{equation} 

where $dA = r^2 \int _{\rho = 0}^{2 \pi} \int _{\iota = - \frac{\pi}{2}}^{  \frac{\pi}{2}}  \cos \iota d \iota d \rho  $ and $ < \cdot > $ implies the time average.

$\frac{dP^{(T)}}{dA} $ is the power emitted in the tensor wave 

\begin{equation} \label{eq:3}
 \frac{dP^{(T)}}{dA}  \equiv \frac{c^3}{16 \pi G(1 - \zeta)} <   \dot{h}^2_{+}(t) + \dot{h}^2_{\times}(t) >  
\end{equation} 

and $\frac{dP^{(S)}}{dA} $ is the power emitted in the scalar wave

 \begin{equation} \label{eq:4}
 \frac{dP^{(S)}}{dA} \equiv \frac{c^3}{16 \pi \zeta G  } <   \dot{h}^2_{s}(t) >
\end{equation}


\subsection{Electromagnetic radiation}

An electric charge at rest produces a static electric field, whereas a charge moving with a uniform velocity gives rise to the magnetic field. Therefore, only an accelerating charge can generate EM radiation. (Note: A charge moving with constant velocity in a vacuum can?t produce the radiation, but it can produce radiation in plasma). So, we can say that there is no monopole radiation in the EM theory because of the charge conservation. The significant contribution in the radiation is due to the time-varying dipole moment, and then there are also quadrupole, octupole moments and so on. We can also understand the property mentioned above in the language of particle physics. A photon is a massless particle with spin 1, so it cannot exist in the state of total angular momentum j = 0, and this implies the absence of the monopole radiation. The two transverse polarizations (vertical and horizontal) in the EM theory differ by 90 degrees, meaning that a rotation of 360 degrees takes us to the same polarization. This can be seen by putting $s = 1$ in Eq. \eqref{eq:0}

The total power emitted in the electromagnetic radiation is given by

 \begin{equation} \label{eq:5}
P_{em} = \frac{2}{3} \frac{K}{c^3} <\ddot{D}_{q}^2 (t')> + \frac{1}{20} \frac{K}{c^5} <\dddot{Q}_{q}^{ij}(t') \dddot{Q}_{q}^{ij}(t')> + \frac{2}{3} \frac{K}{c^3} <\ddot{\mu^2}(t')>
\end{equation} 

where $K = \frac{1}{4 \pi \epsilon_0}$, $D_q (t')$ is the time-varying electric dipole moment in the inertial frame, $Q_{q}^{ij}(t')$ is the time-varying electric quadrupole moment in the inertial frame and $\mu(t')$ is the time-varying magnetic dipole moment in the inertial frame. 


\section{Multipole moments of a SHO}

\subsection{Dipole moment}

Let the position of the particle be $x_p (t) = x_0 \cos (\omega t') $ at an instant $t'$. The x-component of mass dipole moment in the inertial frame can be written as

 \begin{equation} \label{eq:6}
 d(t') = m x(t') = m x_0 \cos (\omega t')
\end{equation} 

The y and z components vanish because the mass distribution is along the x-axis.  The dipole moment in the vector form is

    \begin{equation} 
 \label{eq:7}
D_{m} (t')  =  \begin{bmatrix} d_m(t')   \\ 0  \\ 0  \end{bmatrix}
 \end{equation}  

Consider a wave making an angle $\iota$ with the z-axis. The mass dipole moment in the wave frame can be computed as

 \begin{equation} \label{eq:8}
 D_{mW}(t') = S \cdot D_{m} (t')
\end{equation} 

where $ S $ is the rotation matrix given by

    \begin{equation} 
 \label{eq:9}
S  =  \begin{bmatrix} \cos \iota & 0 & - \sin \iota  \\ 0 & 1  & 0 \\ \sin \iota & 0 & \cos \iota  \end{bmatrix}
 \end{equation}

The matrix multiplication gives

    \begin{equation} 
 \label{eq:10}
D_{mW} (t')  =  \begin{bmatrix} d(t')  \cos \iota  \\ 0  \\ d(t')  \sin \iota  \end{bmatrix}
 \end{equation} 

The analogous charge dipole moment in the inertial frame is

    \begin{equation} 
 \label{eq:11}
D_q (t')  =  \begin{bmatrix} d_q(t')   \\ 0  \\ 0  \end{bmatrix} = \begin{bmatrix} q x_0 \cos (\omega t')   \\ 0  \\ 0  \end{bmatrix}
 \end{equation}

\subsection{Quadrupole moment}

The linear mass density of a point particle oscillating in one dimension can be represented by a Dirac delta function as 
\begin{equation}\label{eq:12}
\lambda (t',x) = m \delta(x-x(t'))
  \end{equation} 

and the expression for the symmetric-trace free (STF) quadrupole moment is 

 \begin{equation} \label{eq:13}
Q_{m}^{ij} (t') = \int \left[ x^i x^j - \frac{1}{3} r^2 \delta _{ij} \right] dm
\end{equation}

The $Q_m^{xx}$ component is

\begin{equation} 
\label{eq:14}
Q_m^{xx}(t') = m \int \delta ( x - x_p (t')   )  dx  \left[ x_p(t')^2  - \frac{1}{3}  \left( x_p(t')^2  + 0  + 0   \right)    \delta_{11} \right] 
 \end{equation} 

or, 

  \begin{equation} 
\label{eq:15}
Q_m^{xx}(t') =  \frac{2}{3} m   \left[  x_{p} (t')  \right]^2  = \frac{2}{3} m x_0^2  \cos^2 (\omega t')
 \end{equation} 

Similarly, we can compute other components of the quadrupole moment tensor to obtain

     \begin{equation}
 \label{eq:16}
Q_{m}^{ij} (t') =  Q_{m0}\begin{bmatrix} 2  & 0 & 0 \\ 0 & - 1  & 0 \\ 0 & 0 & - 1  \end{bmatrix}
 \end{equation} 

where $Q_{m0}$ is defined as

    \begin{equation}
 \label{eq:17}
  Q_{m0}(t') \equiv    \frac{1}{3} m x_0^2 \cos^2 (\omega t') 
 \end{equation}

The following orthogonal rotation gives the mass quadrupole moment tensor in the wave frame

    \begin{equation}
 \label{eq:18}
  Q_{mW}(t') = S \cdot Q_m(t') \cdot S^T 
 \end{equation} 

and we get

      \begin{equation}
 \label{eq:19}
Q_{mW}^{ij}(t')  = Q_{m0}(t') \begin{bmatrix}   2 \cos^2 \iota - \sin^2  \iota & 0 &  3  \sin \iota \cos \iota  \\ 0 & - 1  & 0 \\  3  \sin \iota \cos \iota & 0 &  2 \sin^2 \iota - \cos^2  \iota  \end{bmatrix}
 \end{equation} 

The analogous charge quadrupole moment tensor in the inertial frame is

     \begin{equation}
 \label{eq:20}
Q_{q}^{ij}(t')  =  Q_{q0}(t')\begin{bmatrix} 2  & 0 & 0 \\ 0 & - 1  & 0 \\ 0 & 0 & - 1  \end{bmatrix}
 \end{equation}

where $Q_{q0} (t') $ is defined as

    \begin{equation}
 \label{eq:21}
  Q_{q0}(t') \equiv    \frac{1}{3} q x_0^2 \cos^2 (\omega t') 
 \end{equation}

\section{Power radiated}

The three GW polarizations are obtained using Eq. \eqref{eq:1} and Eq. \eqref{eq:1.1}. Their expressions are

\begin{eqnarray} 
\label{eq:22}
h_{+} (t) &=&  \frac{-2 G}{r c^{2}} (1 - \zeta) m  \frac{v^2}{c^2}  \cos^2 \iota \cos (2 \omega t')
\nonumber \\
h_{\times} (t) &=& 0 
\end{eqnarray}

    \begin{equation}
 \label{eq:22.1}
h_{S}(t) = \frac{2G}{r c^2} \zeta \left[  - m \frac{v}{c} \sin \iota \sin (\omega t') +  \frac{m}{3} \frac{v^2}{c^2} \cos (2 \omega t') \left( 3 \sin^2 \iota - 1 \right)  \right]
 \end{equation}

In the above expressions, we use $v = x_0 \omega$ where $v$ is the maximum speed of the SHO at the mean position.  When $\zeta$ tends to zero, we approach the limit of GR. In this case, the scalar polarization vanishes. Also, there is no cross-polarization from a classic SHO both in BD theory and GR. We have ignored the mass monopole term in the scalar polarization because there is no mass transfer.The radiation due to the dipole moment comes at frequency $\omega$ whereas,  waves due to the quadrupole moment originates at $2 \omega$. 

We use Eq. \eqref{eq:3} and Eq. \eqref{eq:4} to obtain the power emitted in tensor wave and scalar wave respectively. The expressions are

\begin{eqnarray} 
\label{eq:23}
 P^{(T)} &=& \frac{16}{15} \frac{G (1 - \zeta)}{c} m^2 \omega^2   \left( \frac{v}{c}  \right)^4     
\nonumber \\
P^{(S)} &=& \frac{1}{6} \frac{G \zeta}{c} m^2 \omega^2 \left( \frac{v}{c}  \right)^2   + \frac{8}{45} \frac{G \zeta}{c} m^2 \omega^2 \left( \frac{v}{c}  \right)^4  
\end{eqnarray} 

The total power emitted in the gravitational radiation is $P_{grav} =  P^{(T)}  +  P^{(S)}$

    \begin{equation}
 \label{eq:24}
P_{grav} = \frac{16}{15} \frac{G}{c} m^2 \omega^2   \left( \frac{v}{c}  \right)^4 + \frac{1}{6} \frac{G \zeta}{c} m^2 \omega^2 \left( \frac{v}{c}  \right)^2 - \frac{8}{9} \frac{G \zeta}{c} m^2 \omega^2 \left( \frac{v}{c}  \right)^4
 \end{equation}

In this case, we can consider the last two  terms as a correction to GR because of the absence of parameter $\zeta$ in the first term. In practice, we can ignore the energy emitted due to the quadrupole moment because of the extra factors of c in the denominator. 

For the case of electromagnetic radiation, we directly calculate the total power emitted using Eq. \eqref{eq:5} without finding the fields at the point of observation. For the dipole part, we get

    \begin{equation}
 \label{eq:25}
<\ddot{D}_{q}^2 (t')> = < q^2 x_0^2 \omega^4 \sin^2 \omega t ' >
 \end{equation}

and the quadrupole term gives

    \begin{equation}
 \label{eq:26}
<\dddot{Q}_{q}^{ij}(t') \dddot{Q}_{q}^{ij}(t')> = 6  < (\dddot{Q}_{q0})^2  >
 \end{equation}

We ignore the contribution from the magnetic dipole term because it is proportional to the angular momentum. A little algebra yields

    \begin{equation}
 \label{eq:27}
P_{em} = \frac{1}{3} \frac{K}{c} q^2 \omega^2 \left( \frac{v}{c}  \right)^2 + \frac{4}{15} \frac{K}{c} q^2 \omega^2 \left( \frac{v}{c}  \right)^4 
 \end{equation}

To calculate the total power for both gravitational and electromagnetic radiation, we use the fact that $ <\sin^2 n \omega t> = <\cos^2 n \omega t> \approx \frac{1}{2}$  and $ <\sin n \omega t> = <\cos n \omega t> \approx 0$.

The total power radiated by the point particle is $ P_{grav} + P_{em}$. In the absence of any radiation, the total energy of the SHO is constant and equal to $ \frac{1}{2}  m \omega^2 x_0^2 $. But radiation takes away energy and momentum from the system. As a result, the amplitude of oscillations will decrease, and finally, the particle will come to rest. We can find the rate of change of amplitude using the equation

    \begin{equation}
 \label{eq:28}
\frac{d}{dt} \left( \frac{1}{2}  m \omega^2 x_0^2 \right) = P_{grav} + P_{em}
 \end{equation}

\section{Conclusion}

The numerical value of K is $8.98 \times 10^9 N \ m^2 \  C^{-2}$, and that of G is $6.67 \times 10^{-11} N \ m^2 \  kg^{-2}$. Let's consider a 1 Coulomb of charge and 1 kg of mass. It can be easily seen that power emitted in gravitational radiation is too low compared to electromagnetic radiation. That is why we can safely ignore gravitational radiation loss from objects in our vicinity.

\section*{Acknowledgement}

I cordially thank Prof. Andrzej Krolak for his valuable guidance and support during my PhD studies. I also warmly thank Prof. Enrico Barausse for his time and helpful remarks to enhance this article.




\end{document}